\documentclass[pra,twocolumn,aps,superscriptaddress,longbibliography]{revtex4-1}
\usepackage{amsmath}
\usepackage{amssymb}
\usepackage{graphicx}
\usepackage[usenames,dvipsnames]{color}
\usepackage{bm}
\usepackage{times}
\usepackage{hyperref}
\usepackage{float}
\usepackage{lipsum}
\usepackage{braket}
\usepackage[T1]{fontenc}
\usepackage[latin1]{inputenc}
\usepackage[table]{xcolor}
\hypersetup{
  colorlinks=true,
  citecolor=blue,
  linkcolor=blue,
  urlcolor=blue}

\definecolor{orange(colorwheel)}{rgb}{1.0, 0.5, 0.0}
\definecolor{maroon(html/css)}{rgb}{0.5, 0.0, 0.0}
\definecolor{lightgray}{gray}{0.9}

\begin{document}

\title{Enhancement of the Bose glass phase in the presence of an artificial gauge field} 
\author{Sukla Pal}
\affiliation{Physical Research Laboratory,
             Ahmedabad - 380009, Gujarat,
             India}
\affiliation{Department of Physics, Centre for Quantum Science, and Dodd-Walls
            Centre for \\ Photonic and Quantum Technologies, University of
            Otago, Dunedin 9016, New Zealand}
\author{Rukmani Bai}
\affiliation{Physical Research Laboratory,
             Ahmedabad - 380009, Gujarat,
             India}
\affiliation{Indian Institute of Technology Gandhinagar,
             Palaj, Gandhinagar - 382355, Gujarat,
             India}
\author{Soumik Bandyopadhyay}
\affiliation{Physical Research Laboratory,
             Ahmedabad - 380009, Gujarat,
             India}
\affiliation{Indian Institute of Technology Gandhinagar,
             Palaj, Gandhinagar - 382355, Gujarat,
             India}
\author{K. Suthar}
\affiliation{Physical Research Laboratory,
             Ahmedabad - 380009, Gujarat,
             India}
\author{D. Angom}
\affiliation{Physical Research Laboratory,
             Ahmedabad - 380009, Gujarat,
             India}

\begin{abstract}
We examine the effects of an artificial gauge field and finite temperature in a
two-dimensional disordered Bose-Hubbard model. The disorder considered is diagonal 
and quenched in nature. A signature of disorder in the Bose-Hubbard model is the
Bose glass phase. Our work shows that the introduction of an artificial
gauge field enhances the domain of the Bose glass phase in the phase
diagram. Most importantly, the size of the domain can be tuned with the 
strength of the artificial gauge field. The introduction of the finite
temperature effects is essential to relate theoretical results with the 
experimental realizations. For our studies we use the single site and cluster 
Gutzwiller mean-field theories. The results from the latter are more reliable as it better
describes the correlation effects. Our results show that the Bose glass phase has a 
larger domain with the latter method.
\end{abstract}

\maketitle

\section{Introduction}
\label{intro}

The observation of a superfluid (SF) to Mott insulator (MI) transition in an 
optical lattice \cite{greiner_02} have opened a new paradigm to explore the
physics of quantum many-body systems. Optical lattices are clean and highly
controllable; in contrast, the condensed matter systems of interest are never
devoid of impurities. Thus, some of the fundamental questions in condensed 
matter physics are related to quantum phase transitions in the presence of 
disorder. The presence of disorder constrains the evolution of a quantum 
system in the Hilbert space and gives rise to quantum glassy phases 
like Bose glass (BG) \cite{fisher_89,lin_12} and phenomena like Anderson 
localization \cite{anderson_58, schulte_05, billy_08, roati_08}. The early
theoretical investigations of disordered Bose Hubbard model (DBHM) 
\cite{fisher_89, giamarchi_88} showed that there is no MI-SF transition in 
presence of diagonal disorder as the BG phase always occurs as the intermediate
phase. The theorem of inclusion \cite{pollet_09, gurarie_09} agrees well with 
this prediction while identifying BG phase as a Griffiths phase containing 
the rare regions. In these rare-regions, the energy gap of adding 
another boson to the system vanishes and thus can be identified as SF islands.

  The DBHM have been studied with diverse techniques: 
mean field \cite{krutitsky_06}, projected Gutzwiller method \cite{lin_12}, 
site independent and multisite mean-field method 
\cite{buonsante_07, pisarski_11}, stochastic mean field \cite{bissbort_10},
quantum Monte Carlo \cite{gimperlein_05, soyler_11, sengupta_07}, 
density matrix renormalisation group (DMRG) \cite{rapsch_99, gerster_16} for 
1D system and numerous others 
\cite{pai_96, nikolay_04, kruger_09, kruger_11,carrasquilla_10}.
In all the cases the introduction of disorder leads to the emergence of BG 
phase which is characterized by finite compressibility and zero 
superfluid stiffness. In the present work, we study 2D DBHM at finite 
temperatures using single site Gutzwiller and cluster Gutzwiller mean field 
theories. More importantly, we examine the effect of the artificial gauge
fields in DBHM. Here, it must be emphasized that most of the theoretical 
investigations of DBHM are at zero temperatures, but the experimental 
realizations are at finite temperatures. This gap is addressed in the present 
work by examining the consequent effects of thermal fluctuations to the BG 
phase. One key finding is the presence of normal fluid (NF) phase at finite 
temperatures and melting of Bose glass phase. The latter is consistent with 
the findings reported in ref. \cite{thomson_16}. 

  In optical lattices it is possible to create an equivalent of Lorentz force 
with artificial gauge fields
\cite{lin_09,lin_11,dalibard_11,hof_76, garcia_12, aidelsburger_11} and
is associated with what is referred to as synthetic magnetic field.
The introduction of the artificial gauge field breaks time reversal symmetry
and modify the band structure. Through the introduction of tunable artificial
gauge field it has been possible to observe the single particle mobility 
edge \cite{gadway_18} in zig-zag chains. Apart from the transport properties, 
the localization effect of the artificial gauge field can enhance the glassy 
features of DBHM. Indeed, our study reveals that localization in DBHM
can be controlled through the artificial gauge field. For this we use
Edward Anderson order parameter (EAOP) as a measure of localization 
while tuning the strength of artificial gauge field. The EAOP is a measure of 
number fluctuation over disorder realizations and it is finite for the BG 
phase, but zero and close to zero for the MI and SF
phases, respectively. From the values of EAOP we find that
there is enhancement of the BG phase in the presence of artificial gauge
field. From the experimental point of view this is important as it could
facilitate detailed studies of the BG phase.

  Experimentally, DBHM can be realized by the addition of speckle type of 
disorder \cite{clement_05, clement_08, white_09}, or by the generation of 
incommensurate multichromatic lattice \cite{damski_03,fallani_07}. Indirect
measurements on SF-BG transition have been reported in 1D \cite{gadway_11} and 
3D \cite{pasienski_10,meldgin_16} systems through transport and coherence 
measurements. In 2D, the observation of center of mass dynamics \cite{yan_17} 
has been theoretically proposed as a method to detect the BG phase 
while ref. \cite{delande_09} suggests measuring the radius of the atomic cloud.
Replica symmetry breaking \cite{thomson_14, morrison_08} also has been 
proposed as a possible detection scheme. In spite of these various 
proposals and progresses towards the realization of a Bose glass, a clear 
and unambiguous experimental evidence of BG phase is yet to be achieved. 
In future studies, quantum gas microscopes \cite{bakr_09} could probe the
properties of the BG phase as it can study the SF islands in BG phase.
And, recent work has proposed  it as an experimental tool to detect 
BG phases \cite{thomson_16}. 
  
  This paper is organized as follows. In the Section II we give an account of 
the single site and cluster Gutzwiller mean field theories. This is then 
followed by a description of the artificial gauge field and observable 
measures to distinguish different phases in Section III and IV. Then, in 
Section V we provide detailed description of the results obtained from our
studies and discuss our observations. And, we then conclude in Section VI.


\section{Model and Gutzwiller mean field theory}
\label{model}

The DBHM for a square lattice with nearest neighbour hopping is defined by the 
Hamiltonian
\begin{eqnarray}
  \hat{H} &= &-\sum_{p,q}\left [ J_x\left( \hat{b}_{p+1, q}^{\dagger}
              \hat{b}_{p,q} + {\rm H.c.} \right ) 
              + J_y\left( \hat{b}_{p, q+1}^{\dagger}
              \hat{b}_{p,q} + {\rm H.c.}\right ) \right ]
                          \nonumber \\
          && + \sum_{p,q}\hat{n}_{p,q} \left [\frac{U}{2}(\hat{n}_{p,q}
            -1) -\tilde\mu_{p,q} \right],
  \label{dbhm}
\end{eqnarray}
where $p$ ($q$) is the lattice index along $x$ ($y$) axis, 
$\hat{b}_{p,q}^{\dagger}$ ($\hat{b}_{p,q}$) is the creation (annihilation) 
operator for a a boson at the $(p,q)$ lattice site, and $\hat{n}_{p,q}$ is the boson 
density operator; $J_x$ ($J_y$) is the hopping strength between two nearest 
neighbour sites along $x$ ($y$) axis,  $U>0$ is the on-site inter-atomic 
interaction strength, and $\tilde\mu_{p,q} = \mu - \epsilon_{p,q}$ is the local
chemical potential. The disorder is introduced through the random energy 
offset $\epsilon_{p,q}$ which are uniformly distributed independent random 
numbers $r_{p,q} \in [-D, D]$ and $D$ is bound of 
random numbers. Depending on the ratio of $J$ and $U$ the above Hamiltonian 
can describe three possible phases of the system --- MI, BG and 
SF~\cite{fisher_89}. In the strong on-site interaction limit 
$(J/U\rightarrow 0)$ the system is either in the MI phase (gapped phase), or 
in the BG phase. Whereas the system is in SF phase when the tunneling overcomes 
repulsive interaction.


\subsection{Zero temperature Gutzwiller mean-field theory}
\label{gmf}

  In the present work we employ the Gutzwiller mean-field theory to compute the 
properties of the DBHM. In this section we describe two variants of the 
Gutzwiller mean field theory: First is the single site Gutzwiller mean-field 
(SGMF) method, where the lattice sites are correlated through a scalar mean 
field $\phi$ and cannot describe entangled states such as the quantum Hall state. 
And, the second is the cluster Gutzwiller mean field (CGMF) method, which 
incorporates the correlation within a cluster of neighbouring lattice sites 
exactly and inter-cluster correlation through $\phi$. A larger cluster captures the
correlation effects better but at the cost of higher computation.


\subsubsection{SGMF method}
\label{sgmf}

In the SGMF method, $\hat{b}_{p, q}$ ( $\hat{b}^\dagger_{p, q}$) at a 
particular lattice site $(p,q)$ is decomposed into mean field 
$\phi_{p, q}$ ($\phi^{*}_{p, q}$) and fluctuation $\delta \hat{b}_{p, q}$
($\delta \hat{b}^{\dagger}_{p, q}$) parts as
\begin{subequations}
\begin{eqnarray}
  \hat{b}_{p, q}           &=& \phi_{p,q} + \delta \hat{b}_{p, q},       \\
  \hat{b}^{\dagger}_{p, q} &=& \phi^{*}_{p, q} + \delta \hat{b}^{\dagger}_{p, q}
 \label{decompose} 
\end{eqnarray}
\end{subequations}
where, $\phi_{p,q} = \langle\hat{b}_{p,q}\rangle$, and $\phi^{*}_{p, q} = 
\langle\hat{b}^{\dagger}_{p,q}\rangle$ are the mean field and its complex
conjugate, respectively. The expectations are defined with respect to the 
ground state of the system. Employing this decomposition, the Hamiltonian
in Eq. (\ref{dbhm}) is reduced to the SGMF Hamiltonian
\begin{eqnarray}
  \hat{H}^{\rm MF} &=& \sum_{p, q}\Biggr\{-J_x
                        \bigg [ \Big(\hat{b}_{p + 1, q}^{\dagger}\phi_{p, q} 
                        + \phi_{p + 1, q}^{*}\hat{b}_{p, q}  
                        - \phi_{p+1,q}^{*}\phi_{p, q}\Big) 
                   \nonumber\\   
                   && + {\rm H.c.}\bigg ] 
                      - J_y\bigg [ \Big(\hat{b}_{p, q+1}^{\dagger} \phi_{p, q} 
                      + \phi_{p, q+1}^{*}\hat{b}_{p, q} 
                      - \phi_{p, q+1}^{*}\phi_{p, q}\Big)
                   \nonumber\\   
                   && + {\rm H.c.}\bigg ] 
                      + \biggr[\frac{U}{2}\hat{n}_{p, q}
                      (\hat{n}_{p, q}-1) - \tilde{\mu}_{p, q}
                      \hat{n}_{p, q}\biggr] \Bigg \},
\label{mf_hamil}
\end{eqnarray}
where terms up to linear in fluctuation operators are considered and those
quadratic in fluctuation operators are neglected. The total Hamiltonian 
in the above expression can be rewritten as 
$\hat{H}^{\rm MF} = \sum_{p,q}\hat{H}^{\rm MF}_{p,q}$, where 
$\hat{H}^{\rm MF}_{p,q}$ is the single site mean field Hamiltonian. The mean 
field $\phi_{p, q}$ can be identified as the SF order parameter which
defines the MI to BG phase-transition in DBHM. Thus, $\phi_{p, q}$ is
zero, when the  system is in MI phase, and finite in BG as well as in the SF phase. 

To compute the ground state of the system the Hamiltonian matrix of 
$\hat{H}^{\rm MF}_{p,q}$ can be diagonalized for each lattice site $(p, q)$ separately. 
And, then the ground state of the system is direct product of the single
site ground states $\ket{\psi}_{p,q}$. Using the Gutzwiller approximation, the 
ground state of the system is
\begin{eqnarray}
 \ket{\Psi_{\rm GW}} = \prod_{p, q}\ket{\psi}_{p, q}
                       = \prod_{p, q} \sum_{n = 0}^{N_{\rm b}}c^{(p,q)}_n
                         \ket{n}_{p, q},
 \label{gw_state}
\end{eqnarray}
where $N_b$ is the maximum allowed occupation number basis (Fock space basis),
and $c^{(p,q)}_n$ are the coefficients of the occupation number state $\ket{n}$
at the lattice site $(p,q)$. From $\ket{\Psi_{\rm GW}}$  we can calculate
$\phi_{p, q}$, the SF order parameter, as
\begin{equation}
\phi_{p, q} = \langle\Psi_{\rm GW}|\hat{b}_{p, q}|\Psi_{\rm GW}\rangle 
            = \sum_{n = 0}^{N_{\rm b}}\sqrt{n} 
              {c^{(p,q)}_{n-1}}^{*}c^{(p,q)}_{n}.
\label{gw_phi}              
\end{equation}
From the above expression it is evident that $\phi_{p, q}$ is zero in the MI 
phase as only one occupation number state $\ket{n}$ contributes to 
$\ket{\psi}_{p,q}$ and hence only one $c^{(p,q)}_n$ has nonzero value. 
Similarly, the occupancy and number fluctuation at a lattice site are
\begin{eqnarray}
   \langle \hat{n}_{p,q}\rangle &=& 
    \sum_{n = 0}^{N_{\rm b}} | c_n^{(p,q})|^2 n_{p,q},\label{number} \\
      \delta n _{p,q} &=& \sqrt{\langle \hat{n}_{p,q}^2\rangle 
                       - \langle \hat{n}_{p,q}\rangle ^2 }
  \label{deltan} 
\end{eqnarray}
In the MI phase $\delta n _{p,q}$ is zero, which makes MI phase incoherent. In 
the BG and SF phase $\delta n _{p,q}$ has nonzero value, but the value of 
$\delta n _{p,q}$ in the BG phase is very small which arises due to
the presence of SF islands in the BG phase. The nonzero and 
relatively large $\delta n _{p,q}$ in the SF phase indicates strong phase 
coherence. Thus $\delta n _{p,q}$ can also be considered as the
order parameter for MI-BG phase transition.


\subsubsection{CGMF method}
\label{cgmf}
 In the CGMF method, to incorporate the hopping term exactly and hence improve
the correlation effects, the total lattice considered is partitioned into 
clusters. That is, for an optical lattice of dimension $K\times L$, we can 
separate it into $W$ clusters ($C$) of size $M\times N$, that is 
$W=(K\times L)/(M\times N)$. Thus, the case of CGMF with $M = N = 1$ is 
equivalent to SGMF.  In CGMF, the kinetic energy or the hopping term is decomposed
into two types. First is the intra-cluster or hopping within the lattice sites
in a cluster, and second is the inter-cluster which is between neighbouring 
lattice sites which lie on the boundary of different clusters. The details of
the present implementation of the CGMF method is reported in ref. \cite{bai_18} 
and the Hamiltonian of a single cluster is
\begin{eqnarray}
 \hat{H}_C & =& -{\sum_{p, q \in C}}'\biggr[J_x 
              \hat{b}_{p+1, q}^{\dagger}\hat{b}_{p, q} 
              + J_y \hat{b}_{p, q+1}^{\dagger}\hat{b}_{p, q}
              + {\rm H.c.}\biggr]\nonumber\\
              &&-\sum_{p, q\in \delta C}
              \biggr[J_x (\phi^c_{p+1,q})^{\ast}\hat{b}_{p, q} 
              + J_y (\phi^c_{p,q+1})^{\ast}\hat{b}_{p, q}
              + {\rm H.c.}\biggr]\nonumber\\
           && +\sum_{p, q \in C}
              \biggr[\frac{U}{2}\hat{n}_{p, q}(\hat{n}_{p, q}-1) - 
              \tilde{\mu}_{p, q}\hat{n}_{p, q}\biggr] 
\label{cg_hamil}         
\end{eqnarray}
where $(\phi^c_{p,q})^{\ast} = \sum_{p^{'},q^{'}\not\in  C} \langle 
b_{p^{'},q^{'}}\rangle$ is the SF order parameter at the lattice site $(p, q)$ which lies
at the boundary of neighbouring cluster. The prime in the summation of the 
first term is to indicate that the $(p+1,q)$ and $(p,q+1)$ lattice points are
also within the cluster. And, in the second term $\delta C$ denotes the lattice
sites at the boundary of the cluster. The matrix element of $\hat{H}_C$ is
defined in terms of the cluster basis states
\begin{equation}
  \ket{\Phi_c}_\ell = \prod_{q=0}^{N-1}\prod_{p=0}^{M-1} \ket{n_p^q},
\end{equation}
where $\ket{n_p^q}$ is the occupation number basis at the $(p,q)$ lattice
site, and $\ell \equiv \{n_0^0, n_1^0, \ldots, n_{M-1}^0, n_0^1, n_1^1,\ldots
n_{M-1}^1, \ldots, n_{M-1}^{N-1}\}$ is 
the index quantum number to identify the cluster state. After diagonalizing 
the Hamiltonian, we can get the ground state of the cluster as
\begin{equation}
   |\Psi_c\rangle = \sum_{\ell} C_\ell\ket{\Phi_c}_\ell.
\end{equation}
where $C_\ell$ is the coefficient of the cluster state.
The ground state of the entire $K\times L$ lattice, like in SGMF, is the direct
product of the cluster ground states
\begin{equation}
 \ket{\Psi^c_{\rm GW}} = \prod_k\ket{\Psi_c}_k
 \label{cgw_state}
\end{equation}
where, $k$ is the cluster index and varies from 1 to 
$W$. The SF order parameter $\phi$ is computed similar 
to Eq.~(\ref{gw_phi}) as
\begin{equation}
   \phi_{p,q} = \bra{\Psi^c_{\rm GW}}\hat{b}_{p,q}\ket{\Psi^c_{\rm GW}}.
\label{cgw_phi}              
\end{equation}
With respect to cluster basis, the average occupancy and number fluctuation 
of lattice sites in the $k$th cluster are
\begin{eqnarray}
   \langle \hat{n}\rangle _k &=& \frac{\sum_{p,q\in C} \langle\hat{n}_{p,q}
                                 \rangle _k}{MN} \label{cnumber}  \\  
   (\delta n)_k &=& \sqrt{\langle\hat{n}^2\rangle _k - 
                 \langle\hat{n}\rangle^2_k}.
   \label{cdeltan} 
\end{eqnarray}
For the entire lattice, the average density can be defined as the mean of the
average occupancy of the clusters.


\subsection{Finite temperature Gutzwiller mean field theory}
\label{gutz_t}

 To incorporate finite temperature effects we require the entire set of 
eigenvalues and eigenfunctions obtained from the diagonalization of the mean
field Hamiltonian. So, in the case of SGMF we use all the single site 
eigenvectors $\ket{\psi}^l_{p,q}$ and corresponding eigenvalues $E^l_{p,q}$ to
define the single site partition function
\begin{eqnarray}
  Z = \sum_{l=1}^{N_b}e^{-\beta E_l},
  \label{pf}
\end{eqnarray}
where $\beta = 1/k_BT$, $T$ being the temperature of the system. Since the 
energy $E_l$ is scaled with respect to $U$, $T$ is in units of $U/k_{\rm B}$ or 
in other words in the rest of the paper temperature is defined in terms
of the dimensionless unit $k_{\rm B}T/U$. In a similar way, for the CGMF we 
can define the cluster partition function in terms of the eigenfunctions 
$\ket{\Psi_c}^l$ and the corresponding eigenvalues. 
Using the above description, the thermal average of the SF order parameter at 
the $(p,q)$ lattice site is
\begin{equation}
   \langle \phi_{p,q}\rangle = \frac{1}{Z}\sum_{l}{_k^l\bra{\Psi_c}}
                \hat{b}_{p,q} e^{-\beta E_l} \ket{\Psi_c}_k^l,
\label{phi_t}
\end{equation}
where $\langle\ldots\rangle$ is used to represent thermal averaging and 
$\ket{\Psi_c}_k^l$ is the $l$th excited state of the $k$th cluster within 
which the $(p,q)$ lattice site lies. Similarly, the occupancy or the density 
can be computed as 
\begin{equation}
   \langle\langle \hat{n}_{p,q} \rangle\rangle =  \frac{1}{Z}\sum_{l}
                      {_k^l\bra{\Psi_c}}\hat{n}_{p,q}e^{-\beta E_l}
                      \ket{\Psi_c}^l_k,
   \label{number_t}
\end{equation}
where, following the notations in Eq. (\ref{number}) and (\ref{cnumber}), the 
additional $\langle\ldots\rangle$ represents thermal averaging. Once we 
obtain $\langle\langle \hat{n}_{p,q} \rangle\rangle$, the average density or 
occupancy is 
$\langle\rho\rangle=\langle n \rangle 
= \sum_{p,q}\langle\langle \hat{n}_{p,q} \rangle\rangle /(K\times L)$. Then, 
like defined earlier, the number fluctuation is 
\begin{equation}
   \delta n_{p,q} = \sqrt{\langle\langle \hat{n}^2_{p,q}\rangle\rangle
                 -\langle\langle\hat{n}_{p,q}\rangle\rangle^2}.
\end{equation}
A new feature of considering finite temperature effects is, it is possible to 
have vanishing $\langle \phi_{p,q}\rangle$ but with  non-integer 
$\langle\langle \hat{n}_{p,q} \rangle\rangle$. This heralds a new 
phase in the phase diagram and is referred to as the normal fluid (NF). 
Thus, at finite temperatures SF order parameter can act as the order parameter 
for the NF-BG transition. 
Compared to the NF phase, the MI on the other hand has vanishing 
$\langle \phi_{p,q}\rangle$ and integer 
$\langle\langle \hat{n}_{p,q} \rangle\rangle$. So, with vanishing 
$\langle \phi_{p,q}\rangle$  the change from 
integer value to non-integer $\langle\langle \hat{n}_{p,q} \rangle\rangle$
can be identified as MI-NF transition.


\section{Artificial gauge field}
\label{art_gauge}
  Artificial gauge fields \cite{lin_09,lin_11,dalibard_11} engineered through 
optical fields can create synthetic magnetic fields for charge neutral 
ultracold atoms in optical lattices. This generates an equivalent of Lorentz 
force for these atoms, and optical lattice is, then, endowed with properties
analogous to the quantum Hall system. Such a system is an excellent model 
system to study the physics of strongly correlated states like quantum Hall
states and their properties. The same logic also applies to the DBHM.
In the Hamiltonian description, the presence of an artificial gauge 
field induces a complex hopping parameter $J \rightarrow J\exp(i\Phi)$ and
accordingly the SGMF Hamiltonian in Eq.~(\ref{mf_hamil}) is modified to
\begin{eqnarray}
  \hat{H}^{\rm MF} &=& \sum_{p, q}\Biggr\{-J_x e^{i\Phi}
                        \bigg [ \Big(\hat{b}_{p + 1, q}^{\dagger}\phi_{p, q}  
                        + \phi_{p + 1, q}^{*}\hat{b}_{p, q}  
                                  \nonumber\\   
                   && - \phi_{p+1,q}^{*}\phi_{p, q}\Big) 
                      + {\rm H.c.}\bigg ] 
                      - J_y\bigg [ \Big(\hat{b}_{p, q+1}^{\dagger} \phi_{p, q} 
                                  \nonumber\\   
                   && + \phi_{p, q+1}^{*}\hat{b}_{p, q} 
                      - \phi_{p, q+1}^{*}\phi_{p, q}\Big)
                      + {\rm H.c.}\bigg ] 
                                  \nonumber\\   
                   && + \biggr[\frac{U}{2}\hat{n}_{p, q}
                      (\hat{n}_{p, q}-1) - \tilde{\mu}_{p, q}
                      \hat{n}_{p, q}\biggr] \Bigg \},
\label{mf_hamil_gauge}
\end{eqnarray}
where, $\Phi$ is the phase an atom acquires when it traverses around a unit
cell or plaquette. The artificial gauge field is considered in the Landau gauge 
and the phase for hopping along $x$ direction arises via the Peierl's
substitution \cite{hof_76, garcia_12}. The artificial gauge field, then,
creates a staggered synthetic magnetic flux \cite{aidelsburger_11} along 
$y$ direction.  The phase can also be defined in terms of the $\alpha$, the 
flux quanta per plaquette, through the relation $\Phi = 2\pi\alpha q$, and 
the flux quanta is restricted in the domain $0\le \alpha\le 1/2$.  In the present work, we 
examine the properties of bosons in presence of artificial gauge field while 
experiencing a random local chemical potential. Although, the effect of an 
artificial gauge field on BHM is quite well studied, the same is not true of 
DBHM.


\section{Characterization of states}

  Each of the low temperature phases supported by DBHM has special properties 
and this leads to unique combinations of order parameters as signatures of 
each phase. The values of these order parameters also determine
the phase boundaries. In Table.~\ref{table:tab}, we list the order parameters
corresponding to each phase.


\subsection{Superfluid stiffness and compressibility}

Phase coherence is a characteristic property of the SF phase, and it is 
absent in the other phases (MI, NF and BG) supported by DBHM. Thus in the 
SF phase it requires finite amount of energy to alter the phase coherence, or 
in other words, it acquires stiffness towards phase change. This property is 
referred to as the superfluid stiffness $\rho_s$, and hence plays an important
role in determining the phase boundary between BG and SF phase. To 
compute $\rho_s$, a twisted boundary condition (TBC) is imposed on the state. 
If the TBC is applied along the $x$ direction, the hopping term in the 
DBHM is transformed as 
\begin{equation}
\!\!\!\!J_x(\hat{b}_{p+1,q}^{\dagger}\hat{b}_{p,q} + {\rm H.c})\rightarrow 
             J_x(\hat{b}_{p+1,q}^{\dagger}\hat{b}_{p,q}e^{i2\pi\varphi/L} 
                              + {\rm H.c}),
\label{twist}
\end{equation}
where, $\varphi$ is the phase shift or twist applied to the periodic boundary
condition, $L$ is the size of the lattice along $x$ direction, and 
$2\pi\varphi/L$ is phase shift of an atom when it hops between nearest 
neighbours. Accordingly, $\rho_s$ is computed employing the following 
expression \cite{gerster_16}
\begin{equation}
  \rho_s=\frac{L}{8\pi^2}\frac{\partial^2E_0}{\partial\varphi^2}|_{\varphi = 0}.
\label{stiff}
\end{equation}
where $E_0$ is the ground state energy with TBC.
The SF phase is a compressible state as $\delta n$ is finite. However, MI phase
and strongly correlated phase like quantum Hall states are incompressible. 
Thus, the compressibility $\kappa$ is a property of the system which can be
employed a diagnostic to support the phases determined through the order
parameters. By definition, $\kappa$ is given by
\begin{equation}
   \kappa=\frac{\partial\langle\hat{n}\rangle}{\partial\mu}.
   \label{sfactor}
\end{equation}
That is, $\kappa$ is the sensitivity of $n$ to the change of $\mu$.

\begin{figure}
  \includegraphics[width=8cm]{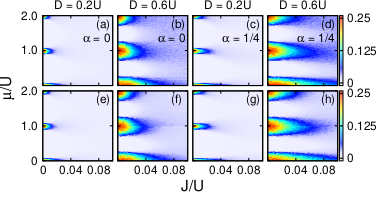}
  \caption{$q_{\text{\tiny{EA}}}$ as a function of $\mu/U$ and $J/U$ at zero 
          temperature. (a)-(d) show $q_{\text{\tiny{EA}}}$ using SGMF method
          and (e)-(h) are obtained employing the CGMF method with 2$\times$ 2
          cluster. (c)-(d) show the enhancement of the BG phase region in the 
          presence of an artificial gauge field with $\alpha = 1/4$ compared to 
          (a)-(b) corresponding to $\alpha = 0$ with disorder strengths
           $D/U= 0.2$ and $0.6$ respectively. This enhancement is also
           captured in (g)-(h) for $\alpha = 1/4$ compared to (e)-(f) using the
           CGMF method.The increase of BG regions with an increase of $D/U$ is
           also notable for both in the presence and in the absence of an artificial
           gauge field. In all the above figures, $q_{\text{\tiny{EA}}}$ is
           obtained by averaging over 50 different disorder distributions. }
    \label{eaop-t0}
\end{figure}

\subsection{Edwards-Anderson order parameter}

For a disordered system the natural and hence, more appropriate order parameter
is the Edwards-Anderson order parameter (EAOP). It can distinguish the 
Griffiths phase by its non zero value and can describe the effect of disorder 
better than other properties like  $\rho_s$, $\kappa$, structure factor, etc.
In the studies with mean field theory, EAOP was first introduced to detect the 
non trivial breaking of ergodicity. Since then various type of EAOP have been 
proposed in literature \cite{morrison_08,graham_09,thomson_14,khellil_16}. In 
our study we consider the EAOP of the following form \cite{thomson_14}
\begin{eqnarray}
   q_{\text{\tiny{EA}}} = \overline{\langle{\hat{n}}_{p,q}\rangle^2}
       -\overline{\langle{\hat{n}}_{p,q}\rangle}^2,
\label{eaop}
\end{eqnarray}  
where, $n_{p,q}$ is the number of atoms at the $(p,q)$ lattice site. The above 
expression involves two types of averages: $\langle\cdots\rangle$ represents
thermal; and $\overline{\cdots}$ indicates average over disorder distribution.
For the $\langle\cdots\rangle$ we consider all the excited states.
From the definition, as the MI phase is identified by integer 
values of $\langle{\hat{n}}_{p,q}\rangle$ $q_{\text{\tiny{EA}}}$ is zero. In 
the SF phase $\langle{\hat{n}}_{p,q}\rangle$ is real and $\delta n_{p,q}$ is 
finite, however, for the clean system $q_{\text{\tiny{EA}}}$ is 
zero as $\langle{\hat{n}}_{p,q}\rangle$ is homogeneous. With disorder, 
$\langle{\hat{n}}_{p,q}\rangle$ is inhomogeneous in the SF phase and hence, 
$q_{\text{\tiny{EA}}}$ is finite but small $O(10^{-3})$ \cite{thomson_16}. In 
the BG phase $q_{\text{\tiny{EA}}}$ is relatively large due to correlation 
between number density and disorder. Thus using $q_{\text{\tiny{EA}}}$ the BG 
phase is distinguishable from MI and NF phases in the present in the system. In 
zero temperature limit we define $q_{\text{\tiny{EA}}}$  as
\begin{eqnarray}
   q_{\text{\tiny{EA}}}|_{(T=0)} = \overline{\langle{\hat{n}}_{p,q}\rangle^2}
       -\overline{\langle{\hat{n}}_{p,q}\rangle}^2,
   \label{eaop0}
\end{eqnarray}  
where we consider expectations only for the ground state.
\begin{table}[h!]
  \begin{ruledtabular}
    \begin{tabular}{lr} 
      \textbf{Quantum phase} & \textbf{Order parameter} \\
      \colrule
      Superfluid (SF) &  $q_{\text{\tiny{EA}}} > 0$, $\rho_s > 0$, $\kappa > 0$, $\phi\ne 0$ \\
      Mott insulator (MI) & $q_{\text{\tiny{EA}}} = 0$, $\rho_s = 0$, $\kappa = 0$, $\phi = 0$ \\
      Bose glass (BG) & $q_{\text{\tiny{EA}}} > 0$, $\rho_s = 0$, $\kappa > 0$, $\phi\ne 0$ \\
      Normal fluid (NF) & $q_{\text{\tiny{EA}}} > 0$, $\rho_s = 0$, $\kappa > 0$, $\phi= 0$\\
    \end{tabular}
    \caption{ Classification of quantum phases and the order parameters 
              supported by DBHM at zero and finite temperatures.}
      \label{table:tab}
  \end{ruledtabular}
\end{table}


\section{Results and Discussions}
\label{results}

  To compute the ground state of the system and determine the phase diagram,
we scale the parameters of the DBHM Hamiltonian with respect to the
interaction strength $U$. So, the relevant parameters of the model are
$J/U$, $\mu/U$ and $D/U$. We, then, determine the phase diagram of the DBHM
in the $J/U-\mu/U$ plane for different values of $D/U$, and one unique feature
of the model is the emergence of the BG phase. The local glassy nature of
the BG phase leads to very different properties from the incompressible and
gapped MI phase, and compressible and gapless SF phase. Thus as mentioned
earlier, one of the key issues in the study of DBHM is to identify appropriate
order parameters to distinguish different phases. And, in particular, to
determine the BG phase without ambiguity based on its local properties. To
construct the phase diagram, we consider a $12\times 12$ square
lattice superimposed with a homogeneous disorder distribution. 

In DBHM, depending on the magnitude of $D/U$, the phase diagrams can be
classified into three broad categories. First, at low disorder strength
$D/U \leqslant 0.1$, BG phase emerge in the phase diagram. Second, at moderate
disorder strengths $0.2\leqslant D/U \leqslant 1$, the domain of BG phase is 
enhanced.  This is the most important regime to explore the physics of BG 
phase. The distinctive features in this range consist of shrinking of MI phase
and enhancement of the BG phase. Finally, at very high disorder strengths 
$D/U > 1$, the MI phase disappears and DBHM supports only two phases, BG and SF.
For reference the selected zero temperature results are shown in the 
Appendix.


\subsection{Zero temperature results}
\label{t0}

  The synthetic magnetic field arising from the introduction of the artificial 
gauge field localizes the bosons and suppresses their itinerant property. 
This manifests as a larger MI lobe in the presence of artificial gauge field.  
However, locally the combined effect of disorder and artificial gauge field 
favours the formation of SF islands. This synergy, then, creates a larger domain 
of BG phase in the phase diagram. In terms of identifying the phase boundaries, 
unlike in the $\alpha=0$ where  $\rho_s$ has linear dependence
on $J/U$ in the SF domain, $\rho_s$ cannot be used here as it exhibits no dependence 
on $J/U$. The two possible causes of this are: the TBC required to compute 
$\rho_s$ modifies the magnetic unit cell associated with the chosen value of 
$\alpha$; and with $\alpha\neq 0$ the SF phase contains vortices which reduce 
the SF phase coherence. So, we use $q_{\text{\tiny{EA}}}$ as the order 
parameter to distinguish BG phase from the MI and SF phases. For consistency 
we compute  $q_{\text{\tiny{EA}}}$ both for $\alpha = 0$ and $\alpha = 1/4$ 
employing SGMF and the results are shown in Fig.~\ref{eaop-t0}(a)-(d), 
where  $q_{\text{\tiny{EA}}}$ is shown as a function of $\mu/U$ and $J/U$. 
The general trend is that  $q_{\text{\tiny{EA}}}$ is 
zero in MI and  O$(10^{-3})$ in the SF phase, and O$(10^{-1})$ in BG phase.  From 
the figure, the presence of the BG phase between different MI lobes is 
discernible from the finite values of $q_{\text{\tiny{EA}}}$ and it is 
consistent with the phase diagram determined from $\rho_s$ shown in 
Fig.~\ref{ph-dia-al0}(g)-(j) in Appendix. We can define sharp MI-BG and SF-BG 
boundaries in the phase diagram by defining a threshold value 
of $q_{\text{\tiny{EA}}}$ between the Mott lobes, however, this is 
non-trivial for the patina of BG phase present at the tip of Mott lobes. 
This is the domain where the MI-SF quantum phase transition is driven by phase 
fluctuations and consequently, the number fluctuation is highly suppressed. As 
a result the value of $q_{\text{\tiny EA}}$ is negligible and it cannot be 
used to distinguish BG and SF phases \cite{buonsante_07,bissbort_10}. Thus, to 
identify the BG domain it is essential to complement the results from 
$q_{\text{\tiny EA}}$ with those of other quantities.

  For $\alpha = 1/4$, the region with finite values of $q_{\text{\tiny{EA}}}$ 
increases significantly. This is discernible from the plot of 
$q_{\text{\tiny{EA}}}$ in Fig.~\ref{eaop-t0}(d). For the case of $D/U = 0.6$, 
when $\alpha = 1/4$, the $q_{\text{\tiny{EA}}}$ is finite with a value 
of $\approx 0.2$ upto $J/U \approx 0.03$. Whereas, with $\alpha=0$ as shown in 
Fig.~\ref{eaop-t0}(b), $q_{\text{\tiny{EA}}}$ has similar value only 
upto $J/U \approx 0.02$. This indicates the enhancement of BG region in the
presence of the artificial gauge field. Employing CGMF method with 
$2 \times 2$ cluster, the values of the $q_{\text{\tiny{EA}}}$ obtained are 
shown in Fig.~\ref{eaop-t0}(e)-(h). One important change is that, 
$q_{\text{\tiny EA}}$ is no longer zero in the MI phase, but it is of 
O$(10^{-6})$. This is due to the presence of particle-hole 
excitations in the cluster states. And, the non-zero value of 
$q_{\text{\tiny EA}}$ is consistent with the results reported in a previous 
work \cite{morrison_08}. The figures show similar trends of artificial 
gauge field induced enhancement of the BG region in the phase diagram. The
increase of BG regions with the increase of $D/U$ is also notable for both 
$\alpha = 0$ and $\alpha = 1/4$. Another observation is that, 
$q_{\text{\tiny{EA}}}$ obtained 
from the CGMF method contains less fluctuations and thus describes the boundary 
of SF-BG transition better compared to the SGMF method. Increasing the cluster 
size CGMF can describe the BG-SF boundary more accurately but at the cost of
much higher computational resources. 
\begin{figure}[H]
	\centering
 \includegraphics[width=7.5cm]{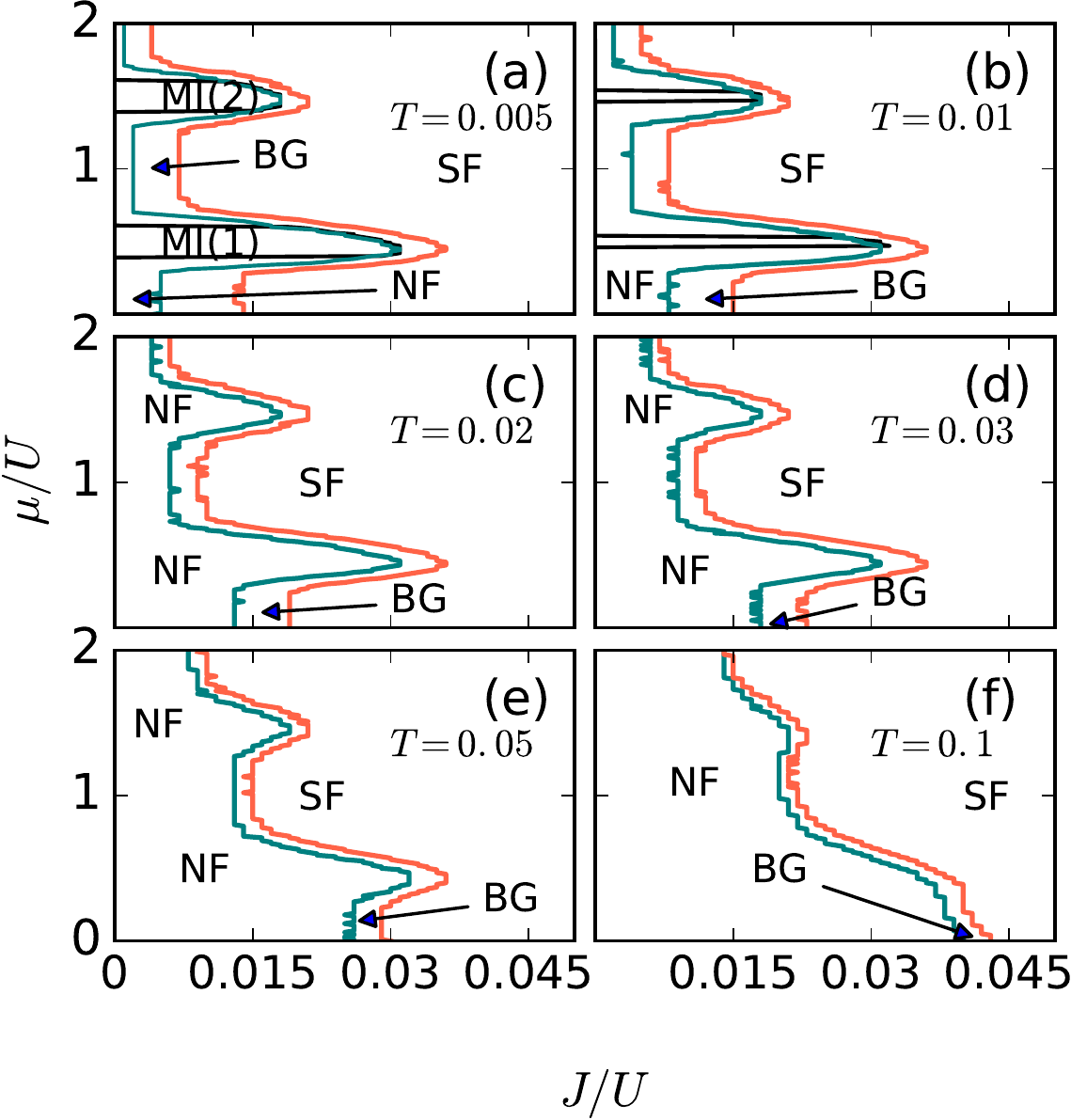}\\
\vskip 0.1cm
 \includegraphics[width=8cm]{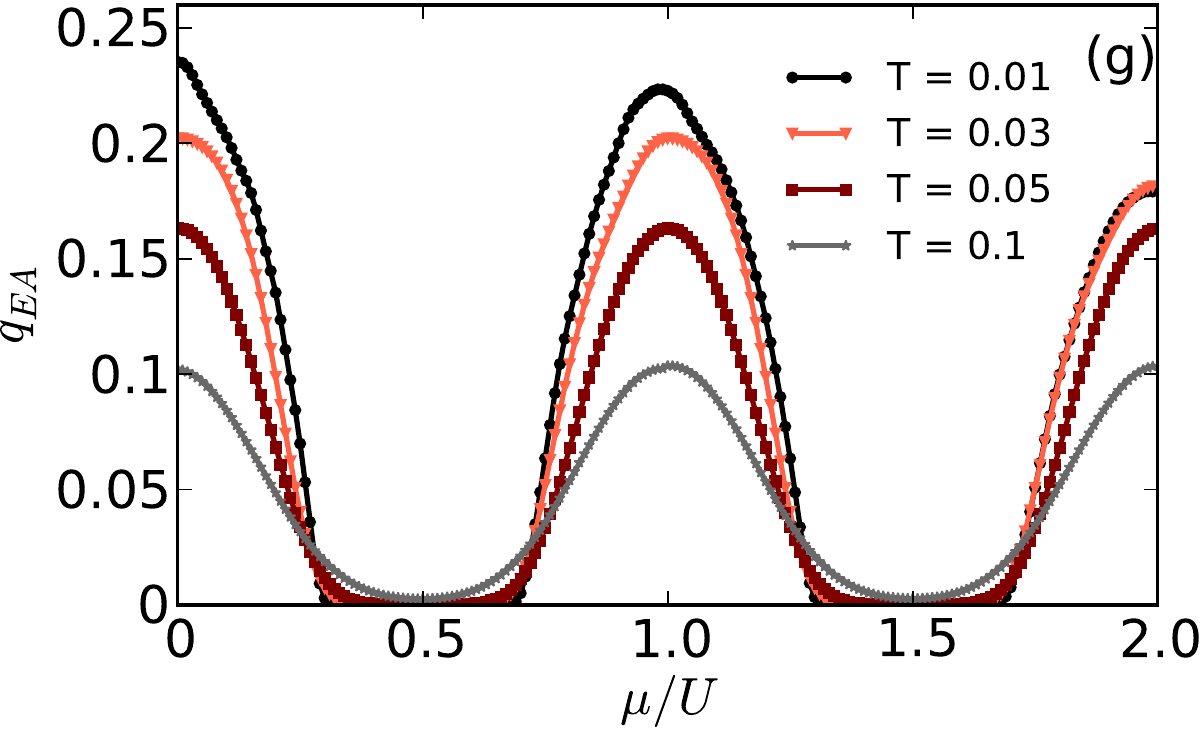}
    \caption{Finite temperature phase diagram using SGMF method
             in absence of artificial gauge
             field for six different temperatures (a) $T = 0.005 U/k_B$, 
             (b)$T = 0.01U/k_B$, (c) $T = 0.02U/k_B$, (d) $T = 0.03U/k_B$,
             (e)$T = 0.05U/k_B$ and (f) $T = 0.1U/k_B$ .
             Disorder strength is fixed at $D = 0.6U$ and each data in the plot
             is obtained by averaging over 500 different disorder
             distributions. (g) shows finite temperature effects on 
             Edward-Anderson order parameter ( $q_{\text{\tiny{EA}}}$) 
             with $D/U = 0.6$ and $J/U$ being fixed at 0.01. The magnitude 
             of $q_{\text{\tiny{EA}}}$ gradually decreases with increase of 
             temperature.}
    \label{ph-dia-t}
\end{figure}


\subsection{Finite temperature results}
\label{ftemp}
 The important outcome of finite temperature is the emergence of a new phase,
the NF phase. This new phase, like the SF phase, has real commensurate 
number of particles per site. But, unlike SF $\phi$ is zero. So, the NF phase 
has some features common to both the MI and SF phases. Previous works reported
the appearance of the NF phase at finite temperatures in the case of the 
canonical Bose-Hubbard model \cite{gerbier_07}, and extended
Bose-Hubbard model with nearest neighbour interactions \cite{ng_10, lin_17}.


\subsubsection{$\alpha=0$}

 The effect of the thermal fluctuations to the $q_{\text{\tiny{EA}}}$, in
absence of artificial gauge field ( $\alpha=0$), is shown in 
Fig.~\ref{ph-dia-t}(g).  The 
results presented in the figure correspond to $D/U=0.6$  and 
each plot is an average over 500 realizations of disorder distributions. With 
increasing temperature there is a monotonic decrease in $q_{\text{\tiny{EA}}}$,
which indicates the {\em melting} of BG phase. Along with the BG phase the MI 
phase also melts, however, this is not apparent from the values of 
$q_{\text{\tiny{EA}}}$. And, the extent of melting can be inferred from the 
phase diagram. To illustrate this point the phase diagram of DBHM at different 
temperatures are shown in Fig~\ref{ph-dia-t}(a-f). As mentioned earlier, 
previous studies have also reported the melting of MI phase due to thermal 
fluctuations \cite{gerbier_07}. But a clear theoretical description and phase 
diagram incorporating finite temperature effects are lacking. Our present work 
shows that the BG phase also melts due to thermal fluctuations. Here, the key 
point is the SF islands, which are hallmark of the BG phase, melts into NF.
This arises from 
the local nature of the SF islands in BG phase, which as a result is affected 
by the local nature of the thermal fluctuations. The bulk SF phase, on the 
other hand, has long range phase correlations and is more robust against 
local fluctuations stemming from finite temperatures. 

\begin{figure}[H]
	\centering
 \includegraphics[width=8cm]{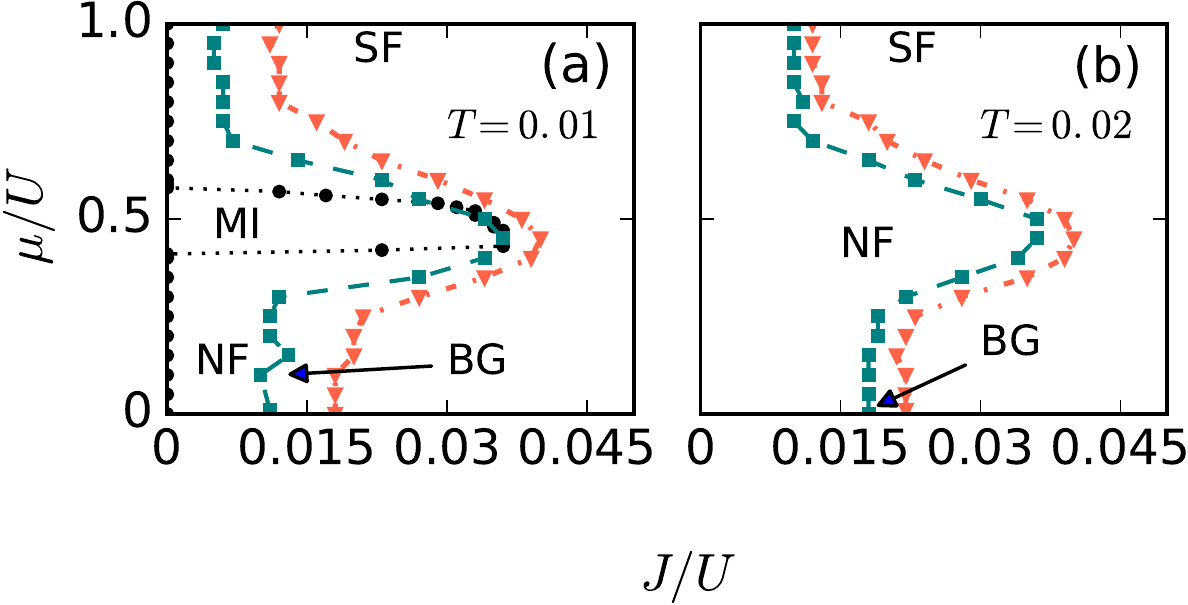}
    \caption{Finite temperature phase diagram using CGMF for $2\times 2$
             cluster in absence of artificial gauge
             field for two different temperatures (a) $T = 0.01 U/k_B$, 
             (b)$T = 0.02U/k_B$;
             Disorder strength is fixed at $D = 0.6U$ and each data in the plot
             is obtained by averaging over 50 different disorder distributions.
             }
    \label{cphd-ft}
\end{figure}

   In the plots the region within the black line is MI phase, whereas, the 
region bounded by the black and green lines is the NF phase, where $\phi$ is 
close to zero  $\phi \leqslant 10^{-6}$. The BG phase lies in the region 
bounded by the green and orange lines, and the area right of the orange line 
is the SF phase. As the temperature is increased, due to the increased thermal 
fluctuations, the phase diagrams undergo several changes. First, the  MI 
lobes shrink and at $k_{\rm B}T/U = 0.02$, MI lobes disappear from the phase 
diagram. This is due to the melting of MI phase and conversion into NF phase. 
So, as discernible from the comparison of Fig.~\ref{ph-dia-t}(a) and (b), the 
MI lobe with $\rho=1$ is bounded and lies in the domain 
$0.40\leqslant \mu/U\leqslant 0.6$ at $k_{\rm B}T/U = 0.005$,
but it shrinks to 0.47 $\leqslant \mu/U\leqslant 0.53 $ at $k_{\rm B}T/U = 0.01$.
Second, the region of the BG phase is reduced with increasing temperature. The 
change is more prominent in the regions which lie between the MI lobes. For 
example, at $\mu=0$ the BG phase exists in the domain 
$0.004\leqslant J/U\leqslant 0.014$ for $k_{\rm B}T/U=0.005$. But, it is 
reduced to  $0.008\leqslant J/U\leqslant 0.015$ when the temperature is 
increased to $k_{\rm B}T/U = 0.01$. As discernible from Fig. ~\ref{ph-dia-t}(f)
at $k_{\rm B}T/U = 0.1$  the domain is reduced to 
$0.04\leqslant J/U\leqslant 0.043$.  And, third, at finite temperatures the 
MI lobes are bounded from top and bottom by straight lines in the SGMF 
results. But, as visible from Fig. \ref{cphd-ft}, the MI boundary is not a 
straight line with CGMF results. This is on account of the better correlation 
effects in CGMF, in contrast, SGMF tends to support sharp NF-MI boundaries 
as a function of $\mu/U$ due to short range coupling through $\phi$.
\begin{figure}[H]
	\centering
 \includegraphics[width=8cm]{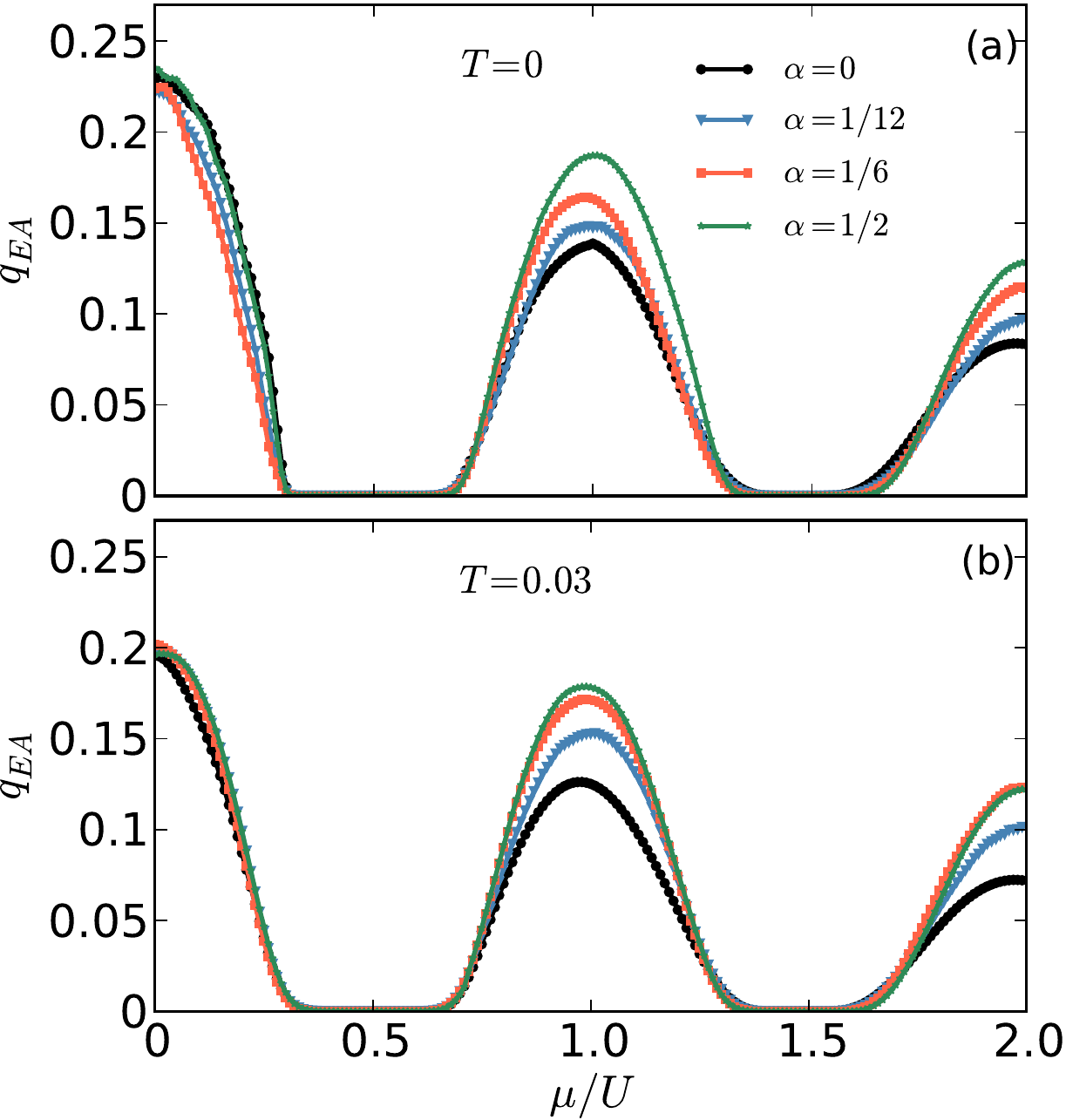}
    \caption{ $q_{\text{\tiny{EA}}}$ as a function of $\mu/U$ for four
             different values of $\alpha$ at (a) $T = 0$ (b) $T = 0.03U/k_B$; 
             with fixed disorder strength $D = 0.6U$ and
             hopping strength $J = 0.02U$.
             In each subfigure  $q_{\text{\tiny{EA}}}$ are calculated for 
             $\alpha = 0, 1/12, 1/6$ and $1/2$
             and averaged over 500 different disorder distributions. 
             }
    \label{eaop-al}
\end{figure}

 Based on the above observations of the phase diagrams at different 
temperatures, the NF-BG and BG-SF phase boundaries shift toward higher 
$J/U$ with increasing temperature. This is due to higher hopping energy 
required to prevail over thermal fluctuations. So that the SF phase is present 
as islands or homogeneous network in BG and SF phases, respectively. The 
other important point is that, the SF phase does not melt directly to NF phase.
In other words, the BG phase advances into the SF phase with ever decreasing 
width with increasing temperature. Thus, the BG phase is an intermediate 
phase between the NF and SF phases. This is the finite temperature equivalent
of the zero temperature phase structure, where BG phase is an intermediate
phase between the MI and SF phases.

 To improve the accuracy of the phase diagram by incorporating additional 
correlation effects, we compute the phase diagram with CGMF using 
$2\times 2$ cluster, and the resulting phase diagram is shown in 
Fig.~\ref{cphd-ft}. The results are for the temperatures $k_{\rm B}T/U = 0.01$ 
and $0.02$, and for better illustration the phase diagrams of only upto 
$\mu/U = 1.0$ are shown in the figure. As to be expected the MI lobes are 
larger in the CGMF results, but the one important change is that the 
envelope of BG phase around the MI and NF phases is more pronounced. 
Consequent to the larger MI lobes, the NF and BG phases encompass
regions with higher $J/U$ compared with the SGMF results. In particular, 
at $\mu =0$ the BG phase occurs in the domain 
$0.011\leqslant J/U \leqslant 0.018$ and $0.018\leqslant J/U \leqslant 0.022$  
for the $k_{\rm B}T/U = 0.01$ and $k_{\rm B}T/U = 0.02$, respectively.


\subsubsection{$\alpha \ne 0$}

\begin{figure}
	\centering
  \includegraphics[width=8cm]{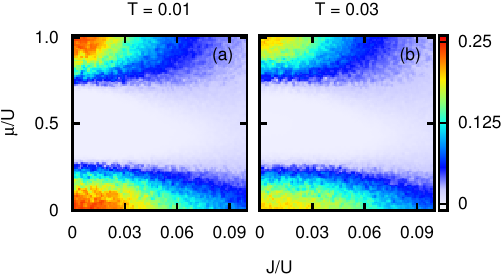}\\
    \caption{ $q_{\text{\tiny{EA}}}$ as a function of $\mu/U$ and $J/U$ for
             $\alpha =1/4$ for two different values
             of temperature $T = 0.01 U/k_B$ (a) and $T = 0.03 U/k_B$. (b) 
             Disorder strength is kept fixed at $D= 0.6 U$ and 
             $q_{\text{\tiny{EA}}}$ are averaged over 50 different disorder 
             distributions with CGMF method. }
    \label{eaop-ft}
\end{figure}

 The thermal fluctuations delocalize the atoms through the entire lattice,
and melt MI phase. This tends to reduce $\phi$. Whereas, as mentioned earlier,
artificial gauge field localizes the atoms, and thereby enhances the
MI lobes. So, these two have opposing effects on the DBHM, and the combined
effects of these two physical factors on the $q_{\text{\tiny{EA}}}$ are shown 
in Fig.~\ref{eaop-al}. In the figure 4 the plots of $q_{\text{\tiny{EA}}}$ for 
$k_{\rm B}T/U =0$ and $0.03$ are shown for different $\alpha$ as a function 
$\mu/U$ at $J/U=0.02$. From the figures it is apparent that the effect of the
artificial gauge field is negligible in the region between the $\rho=0$ and
$\rho=1$ Mott lobes. However, in the regions between other Mott lobes there is 
an enhancement of the BG phase as indicated by the increase in 
$q_{\text{\tiny{EA}}}$. As discernible from Fig.~\ref{eaop-al}(a) the 
value of  $q_{\text{\tiny{EA}}}$ increases from $0.13$ to $0.19$ for the 
region between $\rho=1$ and $\rho=2$ corresponding to 
$0.65 \leqslant \mu/U\leqslant 1.36 $ for non-zero $\alpha$ at 
$k_{\rm B}T/U =0$. From the figure it is also evident that 
$q_{\text{\tiny{EA}}}$ gradually increases with the increase of $\alpha$.
Consequently, the enhancement of BG phase region in DBHM depends on the
strength of artificial gauge field. As a quantitative measure of it, for
$\alpha = 0, 1/12, 1/6$ and $1/2$ $q_{\text{\tiny{EA}}}$
takes the value $0.139$, $0.148$, $0.164$ and $0.187$ respectively around 
$\mu = U$. To demonstrate the combined effect of finite 
temperature and artificial gauge field, the phase diagram in terms of 
$q_{\text{\tiny{EA}}}$ is shown in Fig. \ref{eaop-ft}. As the 
figure is based on $50$ disorder realizations, the general trends of 
$q_{\text{\tiny{EA}}}$ observable in Fig.~\ref{eaop-al} are not apparent. 
However, from the figure the enlargement of the BG phase region between the 
MI lobes is discernible. Thus, this implies that the enhancement of the BG 
phase in presence of artificial gauge field is stable against thermal 
fluctuations.


\section{Conclusions}
\label{conc}
At finite temperatures, the thermal fluctuations lead to melting of the BG 
phase and formation of NF phase. The emergence of the NF phase at finite 
temperatures necessitates using a combination of order parameters and 
properties to identify each phase without ambiguity. More importantly, 
the BG phase is an intermediate phase between the NF and SF phases. 
This is similar to the zero temperature phase where the BG phase is an 
intermediate phase between the MI and SF phases. At higher temperatures the 
melting of MI phase is complete and
only three phases NF, BG and SF phases exist in the system. The addition of 
artificial gauge field brings about a significant change in the phase diagram 
by enhancing the BG phase domain, which is observed in the trends of 
the $q_{\text{\tiny{EA}}}$ without any ambiguity. This implies that such 
enhancements would be observable in quantum gas microscope experiments. To get 
accurate results with mean field theories it is desirable to use the CGMF 
theory. It incorporates correlation effects better and the phase diagrams 
obtained from CGMF are quantitatively different from those obtained from SGMF.


\begin{acknowledgments}

The results presented in the paper are based on the computations
using Vikram-100, the 100TFLOP HPC Cluster at Physical Research Laboratory, 
Ahmedabad, India.

\end{acknowledgments}


\section*{Appendix}

To determine the MI-BG phase boundary, we consider number fluctuation 
($\delta n$) as the property which distinguishes the two phases. In the 
MI phase $\delta n$ is zero for $D/U=0$, however, for $D/U\neq 0$, it is 
non-zero but small due to the disorder. We set $\delta n < 10^{-6}$ as the 
criterion to identify the MI phase in our computations. On the other hand, to 
define the BG-SF boundary, we compute the superfluid stiffness ($\rho_s$). 
In BG phase as the SF phase exists as islands the phase coherence is limited
to these, so the $\rho_s$ small, and we consider $\rho_s < 10^{-2}$ as the 
threshold to distinguish the BG from SF phase. In the SF phase as there is 
phase coherence throughout the system $\rho_s$ is large and it is $O(1)$.  

\begin{figure}[H]
~ \includegraphics[width=7.8cm]{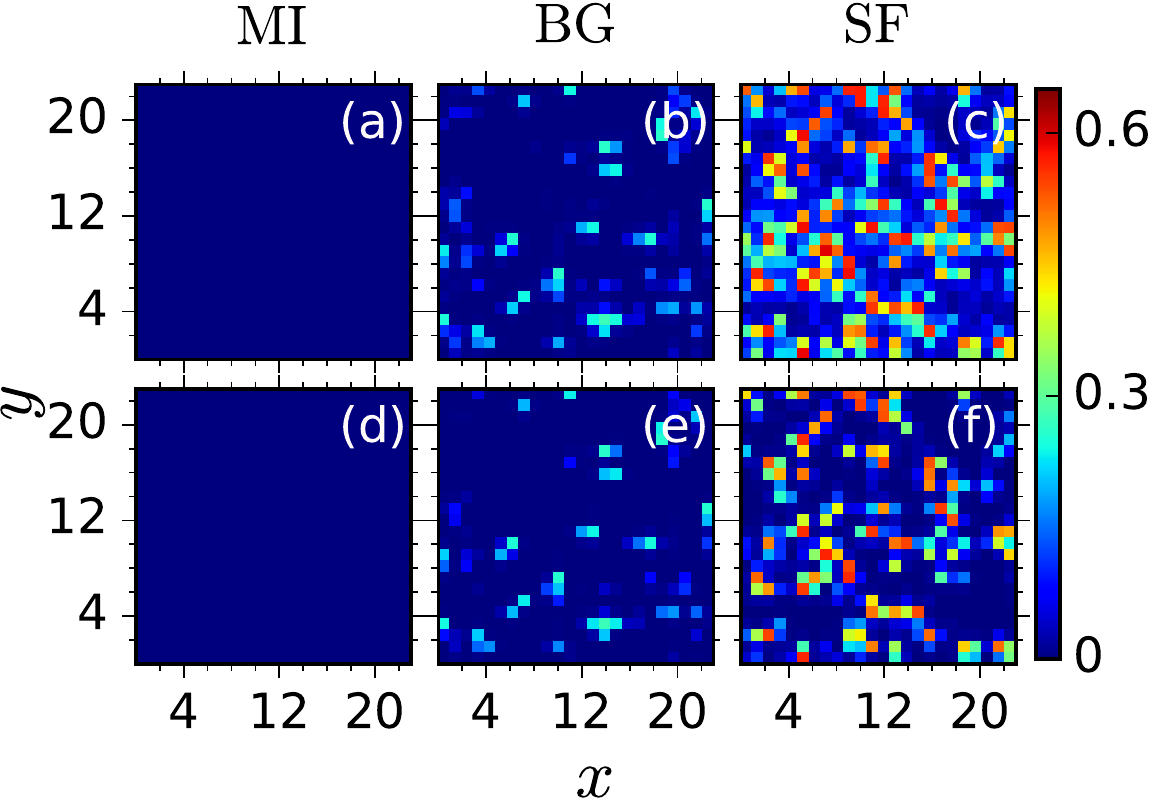}\\
  \includegraphics[width=7.5cm]{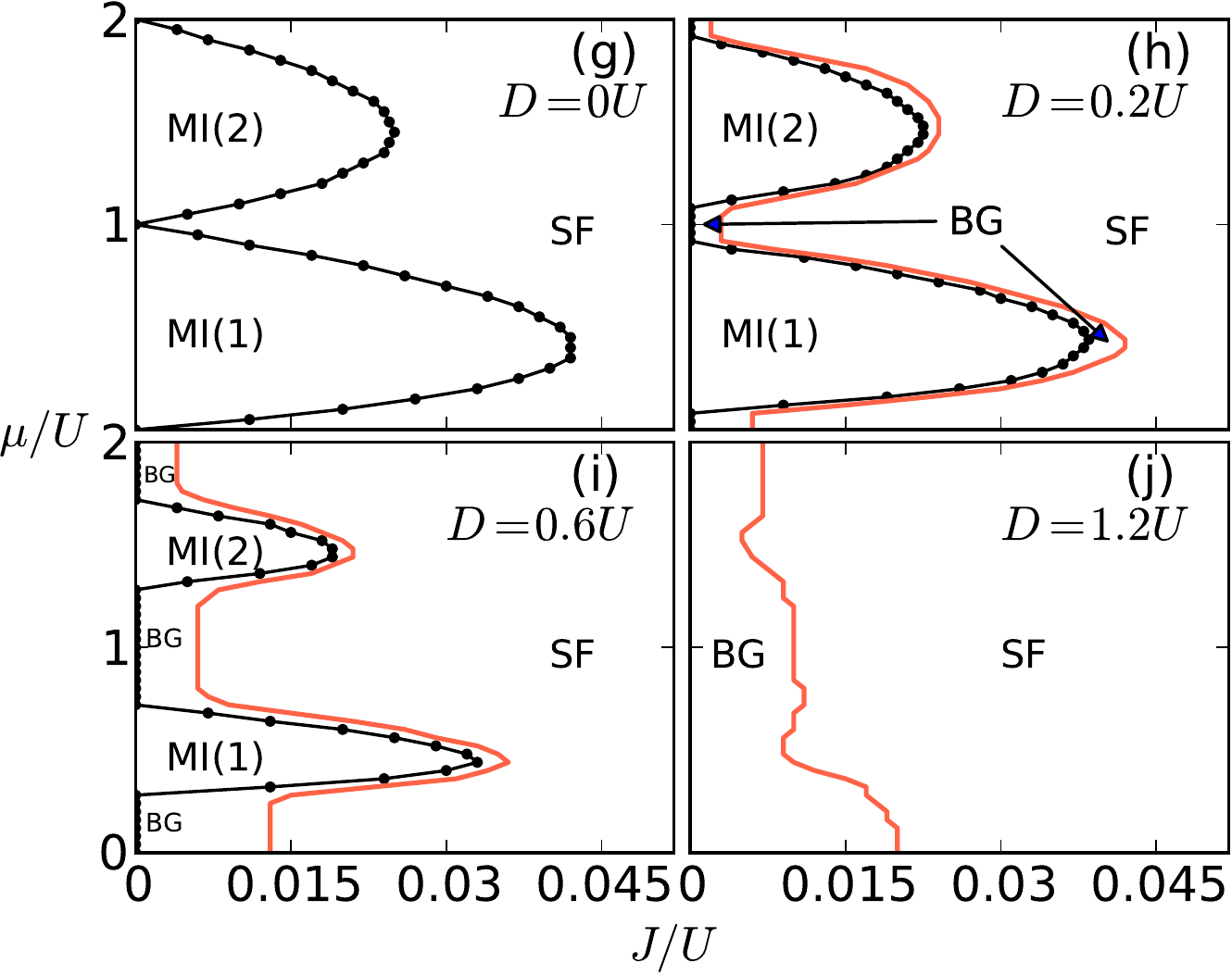}
    \caption{ Order parameter $\phi$ of DBHM at zero temperature for 
              $J/U = 0.01$ and $D/U$ keeping fixed at 0.6 (a)-(c) without and 
             (d)-(f) 
              with ($\alpha=1/4$) artificial gauge field. (a) \& (d) MI phase 
              with $\mu/U = 0.5$; (b) \& (e) BG phase with  $\mu/U = 0.1$; 
              and  (c) \& (f)SF phase with $\mu/U = 1.0$. 
              (g)-(j) equilibrium phase diagram of DBHM using SGMF method at 
              zero temperature in absence of artificial gauge field
              ($\alpha=0$) for disorder strengths $D/U=0$, $0.2$, $0.6$ and
              $1.2$, respectively for 500 different disorder realizations.
              }
    \label{ph-dia-al0}
\end{figure}

 The phase diagrams of DBHM with $\alpha=0$ at  different values of $D/U$ have 
distinctive features \cite{lin_12}. As examples, the phase 
diagrams for the case of $D/U = 0$, $0.2$, $0.6$ and $1.2$ obtained from the
SGMF method are shown in Fig.~\ref{ph-dia-al0}(g)-(j).  With $D/U=0$, the phase 
diagram as shown in Fig.~\ref{ph-dia-al0}(g) consists of only two phases MI 
and SF. With non-zero $D/U$ BG appears in the phase diagram, and as shown in 
Fig.~\ref{ph-dia-al0}(h) for $D = 0.2$ the domain of the MI phase shrinks and an
envelope of BG phase emerges around the MI lobes. From 
Fig.~\ref{ph-dia-al0}(h), it is clear that the BG phase is most prominent in
between the MI lobes.  These are the domains with large density 
fluctuations and small disorder is sufficient to render the bosons itinerant 
to create islands of SF phase. This, then, leads to the formation of BG phase. 
When the $D/U$ is increased to $0.6$, as shown in Fig.~\ref{ph-dia-al0}(i), 
the MI lobes shrink further and the area of the BG phase is enlarged. At 
sufficiently high disorder strength, $D = 1.2U$, the MI phase disappears and 
phase diagram Fig.~\ref{ph-dia-al0}(j) is composed of only SF and BG phases.

\begin{figure}[H]
 \includegraphics[width=8cm]{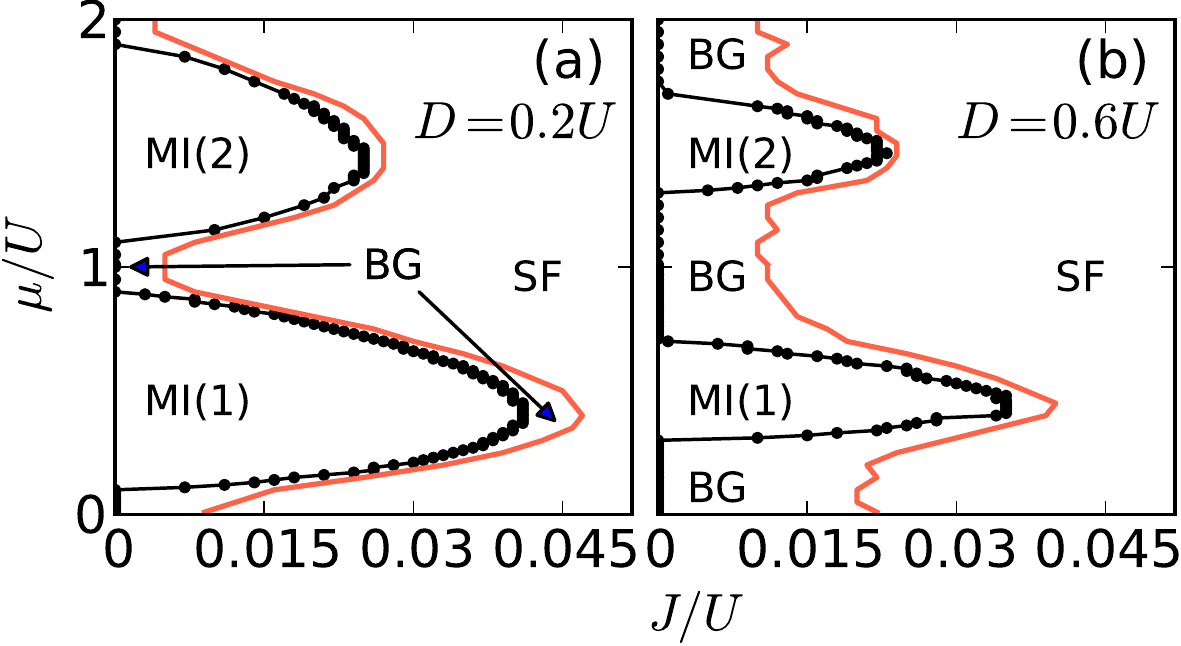}
    \caption{Equilibrium phase diagram of DBHM using CGMF method with cluster
             size 2$\times$2 at zero temperature in absence of artificial
             gauge field for disorder strength (a) $D/U = 0.2$, and  
             (b) $D/U = 0.6$ for 50 different disorder realizations.}
    \label{cph-dia-al0}
\end{figure}

There is an improvement in the phase diagram, which is apparent from the 
enlarged MI lobes, when the phase diagram is computed using CGMF. In particular,
we consider $2\times 2$ cluster and the phase diagrams so obtained are
shown in  Fig.~\ref{cph-dia-al0}. The overall structure of the phase diagram
is qualitatively similar to the SGMF case. However, there are few quantitative
changes. For comparison, consider the case of $D/U = 0.6$, based on our 
results and as visible in Fig.~\ref{ph-dia-al0}(i) and 
Fig.~\ref{cph-dia-al0}(b), there are three important difference due to 
better correlation effects encapsulated in the CGMF method. First, the tip of 
the Mott lobe $\rho=1$ extends upto $0.035$ while it was $0.032$ with SGMF.  
Second, at $\mu/U \simeq 0$, the SF-BG transition occurs at $J/U\approx  0.022$, 
which in the case of SGMF is at $ J/U\approx 0.014$. This is due to the 
association of BG phase with islands of SF phase, and CGMF captures the phase 
correlations in these islands better. The SGMF, on the other hand, tends to 
favour long range phase correlations through the $\phi$ coupling between the 
lattices sites. And, finally, around the tip of the Mott lobes, the area of BG 
phase increases in CGMF method.  

\bibliography{ref}{}

\end{document}